\documentclass[iop,apj,tighten]{emulateapj}
\usepackage{apjfonts} 
\usepackage{amsmath,amstext}
\usepackage[breaklinks,colorlinks,citecolor=blue,linkcolor=magenta]{hyperref} 
\usepackage[all]{hypcap} 

\def\red#1 {\textcolor{red}{#1}\ }   
\def\blue#1 {\textcolor{blue}{#1}\ }   

\shorttitle{PULSED ACCRETION ONTO BINARIES}
\shortauthors{Mu\~noz \& Lai}

\begin{document}

\title{Pulsed accretion onto eccentric and circular binaries}
\author{Diego J. Mu\~noz and Dong Lai}
\affil{Cornell Center for Astrophysics and Planetary Science, 
Department of Astronomy, Cornell University, Ithaca, NY 14853, USA}

\begin{abstract}
We present numerical simulations of circumbinary accretion onto
eccentric and circular binaries using the moving-mesh code
{\footnotesize AREPO}.  This is the first set of simulations to tackle
the problem of binary accretion using a finite-volume scheme on
a freely moving mesh, which allows for accurate measurements of
accretion onto individual stars for arbitrary binary eccentricity.
While accretion onto a circular binary shows 
bursts with period of ${\sim}5$ times the binary period $P_b$,
accretion onto an eccentric binary is predominantly modulated at 
the period ${\sim}1P_b$. For an equal-mass circular binary,
the accretion rates onto individual stars are quite similar to each other,
following the same variable pattern in time.
By contrast, for eccentric binaries, one of the stars
can accrete at a rate 10-20 times larger than its companion.
This ``symmetry breaking" between the stars, however, alternates over timescales of
order 200$P_b$, and can be attributed to a slowly precessing,
eccentric circumbinary disk.  Over longer timescales, the net
accretion rates onto individual stars are the same, reaching a
quasi-steady state with the circumbinary disk.
These results have important implications for the accretion behavior
of binary T-Tauri stars and supermassive binary black holes.
\end{abstract}

\maketitle

\section{Introduction}
Spectroscopic T-Tauri star binaries can exhibit quasi-periodic
photometric oscillations known as ``pulsed accretion'' \citep{jen07,muz13,bary14}.
This variability is believed to arise from the complex accretion streams
delivered onto the young stars from a tidally truncated circumbinary
disk \citep[e.g.][]{art96}. Similar circumbinary disks may also exist around supermassive binary black holes (SMBBHs), 
but the short periods of binary T-Tauri stars (BTTSs) offer
an unparalleled coverage of the time-domain as the system
can evolve over several orbits during observations and 
even {up to hundreds of orbits} between different observing campaigns.
This makes BTTS ideal laboratories for circumbinary accretion 
physics, with direct implications for binary
star and circumbinary planet formation, and
with applications that extend to SMBBHs.

The complexity of circumbinary accretion flow requires direct hydrodynamical
simulations. Several computational approaches have been taken
to address this problem \citep{art96,gun02,mac08,cua09,han10,val11,roe12,shi12,dor13,pel13,far14,line15},
ranging from Lagrangian methods to Eulerian ones on polar and cartesian grids. Of these, only
a subset has been able to simulate the accretion flow onto the individual stars,
or to include eccentricity in the binaries; only two studies have attempted both \citep{gun02,val11}.

Considering eccentric binaries in simulations is essential for pulsed accretion, as 
accretion luminosity is likely to depend on the orbital phase \citep{bas97,hue05,jen07,bary08}.
Although high eccentricities in BTTSs are common -- the binaries AK Sco, DQ Tau and UZ Tau E
have eccentricities of $e_b{=}0.47$, 0.56 and 0.29 respectively \citep{and89,mat97,pra02} -- accurate simulation of
circumbinary accretion onto eccentric pairs remains a challenge.
In this work, we present the first simulation results of circumbinary accretion using the moving-mesh code 
{\footnotesize AREPO} \citep{spr10a}. Unlike other implementations of finite-volume or finite-difference
schemes for computational gas dynamics, the accuracy of {\footnotesize AREPO} does not depend on the value of $e_b$,
as its space-discretization strategy is carried out via an unstructured mesh that moves with the local velocity of the flow.
Thus, being a quasi-Lagrangian method, {\footnotesize AREPO} can naturally concentrate the resolution around the individual stars,
resolving circum-single disks to good accuracy.

\section{Numerical methods}\label{sec:methods}

\subsection{Moving-mesh Hydrodynamics}\label{sec:arepo}
We run two-dimensional, non-selfgravitating hydrodynamic
simulations of viscous circumbinary accretion disks (CBD)
using {\footnotesize AREPO} with a time-explicit
integration scheme for the Navier-Stokes terms \citep{mun13a}. 
The computational domain is divided into
Voronoi cells, distributed in a quasi-polar fashion (with 600 azimuthal zones) with logarithmic
spacing in radius, following the accretion disk setup of \citet{mun14}. Cells initially cover the radial range between $R=a_b(1+e_b)$ to
$R=R_\mathrm{out}=70a_b$ (where $a_b$ is the binary semimajor axis) but are allowed to viscously evolve toward $R=0$.
At $R_\mathrm{out}$, we impose inflow boundary conditions of steady
accretion $\dot{M}_0$, assuming that at these distances the disk 
is axisymmetric and the central potential Keplerian\footnote{
Radial inflow and axisymmetry are imposed by using a moving boundary 
and a narrow wave-absorbing region (as described in \citealp{mun14}) between $65a_b$ 
and $70a_b$.}. 
The binary is represented by a prescribed rotating potential:
\begin{equation}\label{eq:potential}
\Phi(\mathbf{r}) = -\mathcal{G}M_b\left[\frac{~~(1+q_b)^{-1}}{|\mathbf{r}-\mathbf{r}_1|}+
\frac{~q_b(1+q_b)^{-1}}{|\mathbf{r}-\mathbf{r}_2|}\right]~~~~\\
\end{equation}
where $q_b=M_2/M_1$ is the binary mass ratio and $M_b=M_1+M_2$ is the total mass. The
individual stellar positions are $\mathbf{r}_1(t)={q}{(1+q)^{-1}}\mathbf{r}(t)$ and
 $\mathbf{r}_2(t)=-({1+q})^{-1}\mathbf{r}(t)$
where  the relative position vector
$\mathbf{r}(t)=a_b\big(\cos E -e_b~,({1-e_b^2})^{1/2}\sin E\big)$. The
eccentric anomaly $E(t)$ is obtained by solving Kepler's equation \citep[e.g.][]{dan88}. 
The potential around each star is softened\footnote{
We use a cubic spline softening, as in \citet{spr01}.}; 
the softening length is set to $s=0.025a_b$. 

\begin{figure*}[ht!]
\centering
\includegraphics[width=0.91\textwidth]{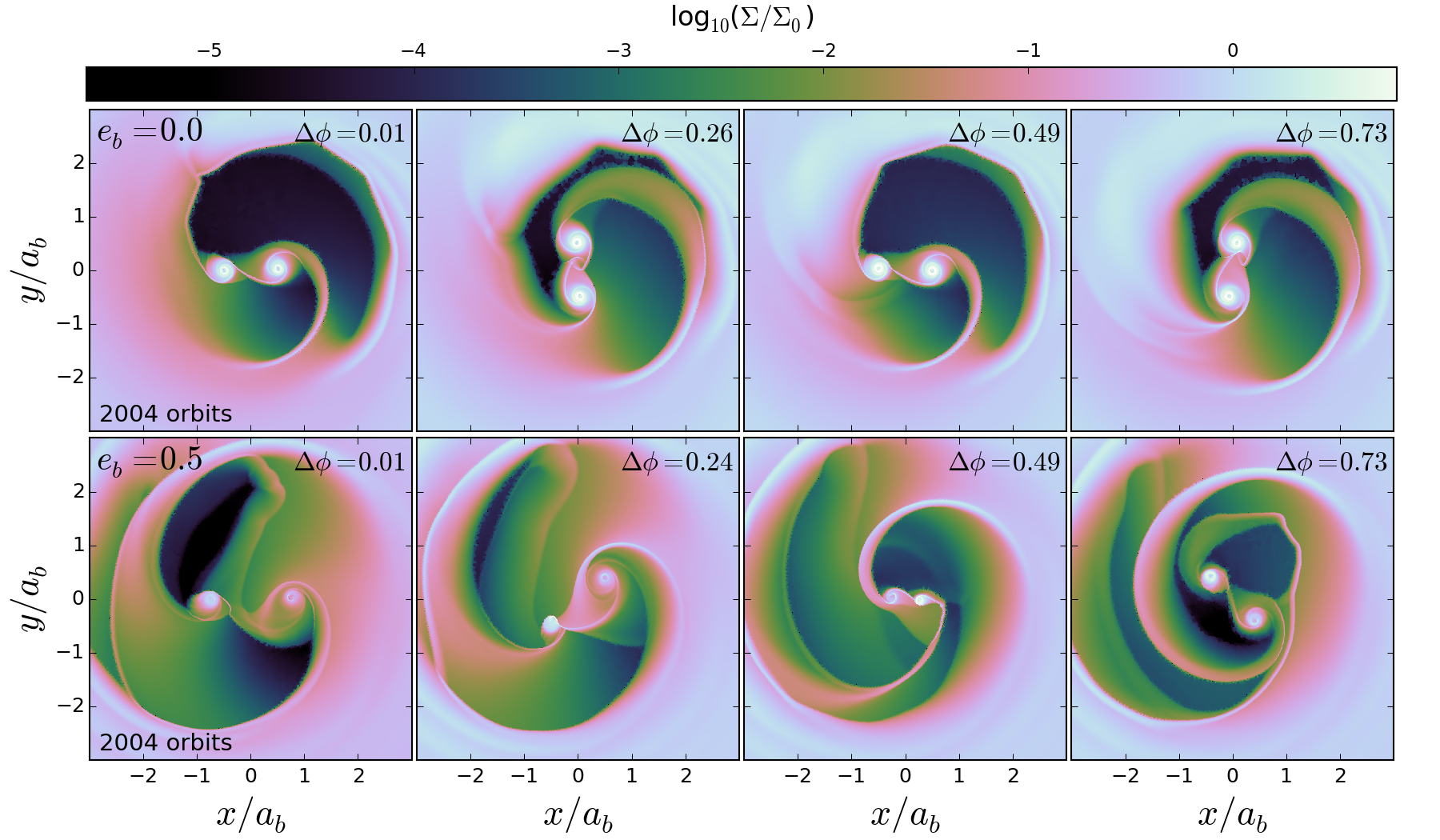}
\caption{Surface density field evolution within timescales of ${\sim}1$ binary orbit for circular (top panels) and eccentric (bottom panels) binaries
at 2004 + $\Delta\phi$ binary orbits (where the relative phase $\Delta\phi\approx0, 0.25, 0.5, 0.75$ is measured since the last apocenter). 
For the $e_b=0$ case, the pattern repeats every half orbit. 
By contrast, in the $e_b{=}0.5$ case, there is a noticeable
asymmetry in the density field. Since $q_b=1$, the asymmetry cannot depend
on the binary properties, but must be imposed by the way gas is funneled into the central cavity. An eccentric inner
disk could favor one star over the other. Note that the accretion rate asymmetry is reversed after several binary orbits (see Section~\ref{sec:individual_accretion})
and Fig.~\ref{fig:accretion_rates_individual}).
\label{fig:cavity_density}}
\end{figure*}

The binary components are allowed to ``accrete''  (although their dynamical
masses are held constant). 
Gas is drained from cells located within a distance of $r_\mathrm{acc}=0.8s=0.02a_b$ from each star. The draining is carried
out as a simple ``open-boundary'' condition, meaning that cells that are located within the accretion
region are instantaneously drained\footnote{
At each time-step, we extract some fraction 
$\eta$ of the cell mass, where $\eta$ is a weighting factor that is unity at $|\mathbf{r}-\mathbf{r}_i|=0$
and decreases with distance from the accreting object.} \citep[][]{mun15}. 
The accreted mass $M_i$
is stored at every major time-step and accretion rates $\dot{M}_1$ and $\dot{M}_2$ are computed by central-difference differentiation of $M_i(t)$.
Note that, for BTTSs with semi-major axis $a_b\sim 0.2$~AU \citep{jen07}, we have $r_\mathrm{acc}\sim0.004~$AU$<1R_\odot$,
sufficient to resolve the accretion down to the stellar surface.
On the other hand, for a SMBBH,  the ``true'' accretion radius (e.g., the innermost stable circular orbit) is $\ll r_\mathrm{acc}$,
and the value of $\dot{M}_i$ should be interpreted with caution (see Section~\ref{sec:resolution} below).

As the outer CBD evolves, resolution is maintained roughly constant via de-refinement and refinement operations \citep{spr10a}.
Within $R=a_b(1+e_b)$, the resolution criterion is switched over from "volume-based" to "mass-based", in which
there is a ``target mass'' $m_\mathrm{gas}$ enforced for all cells \citep{spr10a}. The transition
between mass-based and volume-based resolution is kept smooth by controling  the volume difference between contiguous cells \citep{pak13b}.
For our lowest-resolution runs,
$m_\mathrm{gas}=6.3\times10^{-7}\Sigma_0 a_b^2$, where $\Sigma_0$ is the scaling of the
initial disk surface density profile (see Section~\ref{sec:initial} below). In this region, there is a minimum permitted volume,
$\pi s^2/20$, where $s$ is the softening parameter.

The equation of state is ``locally isothermal''\footnote{
This equation of state requires an isothermal (iterative or approximate) Riemann solver with
varying sound speed computed at each cell interface \citep{mun14}.}: 
$P=\Sigma c_s(\mathbf{r})$,
where the sound speed is a function of position only \citep[e.g.][]{far14}:
$c_s^2(\mathbf{r})=-h_0^2\Phi(\mathbf{r})$,
where the aspect ratio $h_0$ is a global constant.
When $|\mathbf{r}|\gg |\mathbf{r}_1|,|\mathbf{r}_2|$, then $c_s^2\approx h_{0}^2~{\mathcal{G}M_b}/{|\mathbf{r}|}$;
and when $|\mathbf{r}-\mathbf{r}_i|\ll  |\mathbf{r}-\mathbf{r}_j|$, then $c_s^2\approx h_{0}^2~{\mathcal{G}M_i}/{|\mathbf{r}-\mathbf{r}_i|}$.

Finally, the kinematic viscosity $\nu$ follows an $\alpha$-viscosity prescription \citep{sha73}, in which
$\nu=\alpha c_s^2/\tilde{\Omega}(\mathbf{r})$. Where
$\tilde{\Omega}(\mathbf{r})$ is a function that reduces to $({\mathcal{G}M_b/{R}^3})^{1/2}$ far from the
binary, and to $({\mathcal{G}M_b/{|\mathbf{r}-\mathbf{r_i}|^3}})^{1/2}$ close to each star.

Throughout this work, we fix the parameters
$q_b=1$ and $h_0=\alpha=0.1$, while varying the binary eccentricity $e_b$.

\subsection{Initial Setup}\label{sec:initial}
Knowing that the outer disk is in steady-state accretion,
we ``guess'' an initial surface density profile $\Sigma(R)$ that includes a central cavity but that, at
large radii, behaves as $\Sigma\propto \dot{M}_0/\nu\propto R^{-1/2}$. Thus, we adopt the initial CBD surface density profile:
\begin{equation}\label{eq:surface_density}
\Sigma(R,t{=}0)=\Sigma_0\left(\frac{R}{R_{\mathrm{cav},0}}\right)^{-p}
\exp\left[-\left(\frac{R}{R_{\mathrm{cav},0}}\right)^{-\xi}\right]~~,
\end{equation}
where $p=1/2$ and $R_{\mathrm{cav},0}$ and $\xi$ characterize the extent and steepness
of the tidal cavity around the binary. In this work, we choose $R_{\mathrm{cav},0}=5a_b$ and $\xi=4$.
Steady-state at the onset of the simulation\footnote{
Imposing $\Sigma\propto \dot{M}_0/\nu\propto R^{-1/2}$ in the outskirts of the disk is essential
for a steady-state solution to exist, and the only way to
meaningfully compare binary accretion rates to those of single point masses (see
\citealp{raf16} for a similar argument).} 
 is  guaranteed for $R\gg R_{\mathrm{cav},0}$ by construction, but
Eq.~\ref{eq:surface_density} is still an imperfect initial condition at intermediate radii, 
and a long integration time may be needed to relax
the initial conditions for {\it all} $R$.

\begin{figure*}[ht!]
\centering
\includegraphics[width=0.4\textwidth]{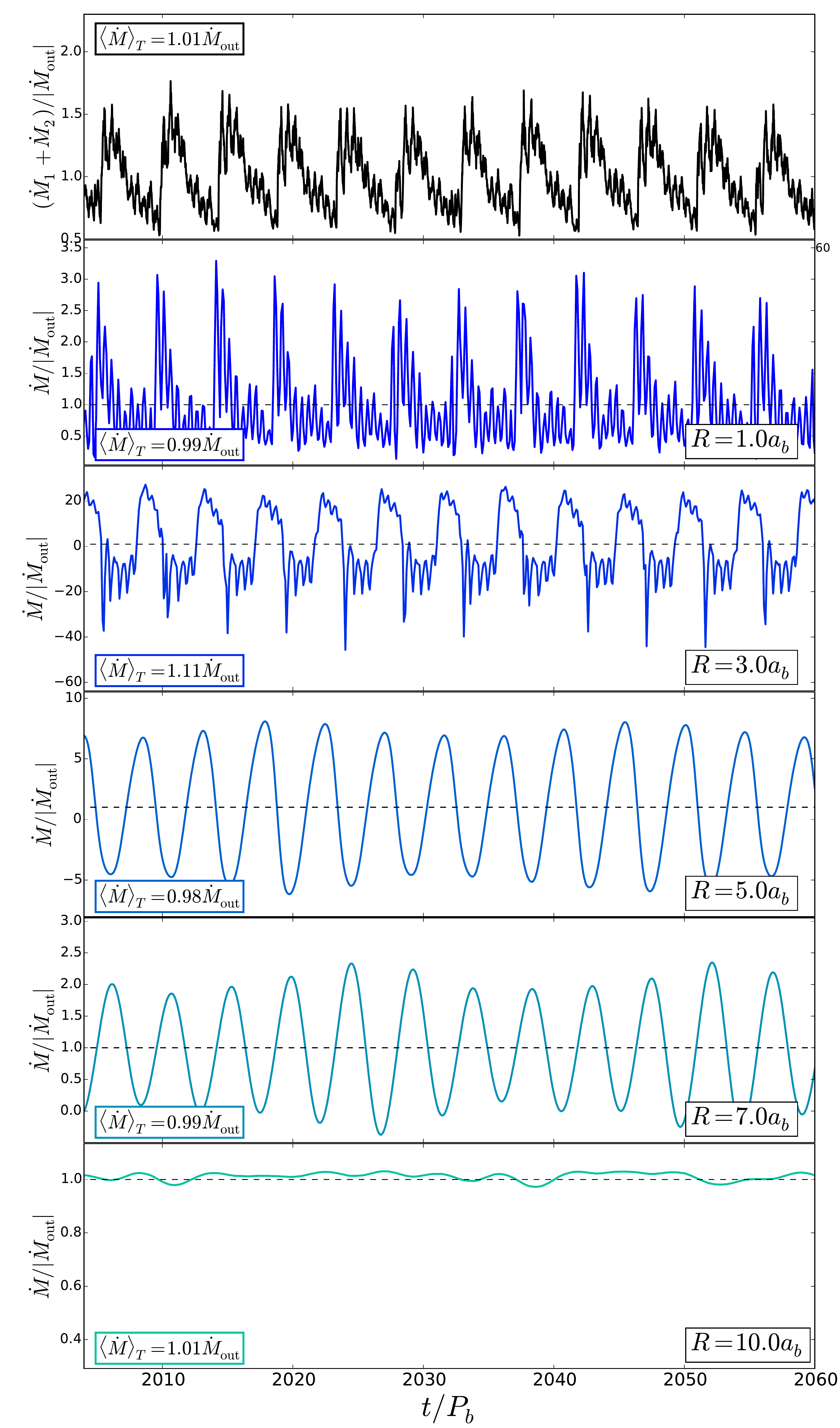}
\includegraphics[width=0.4\textwidth]{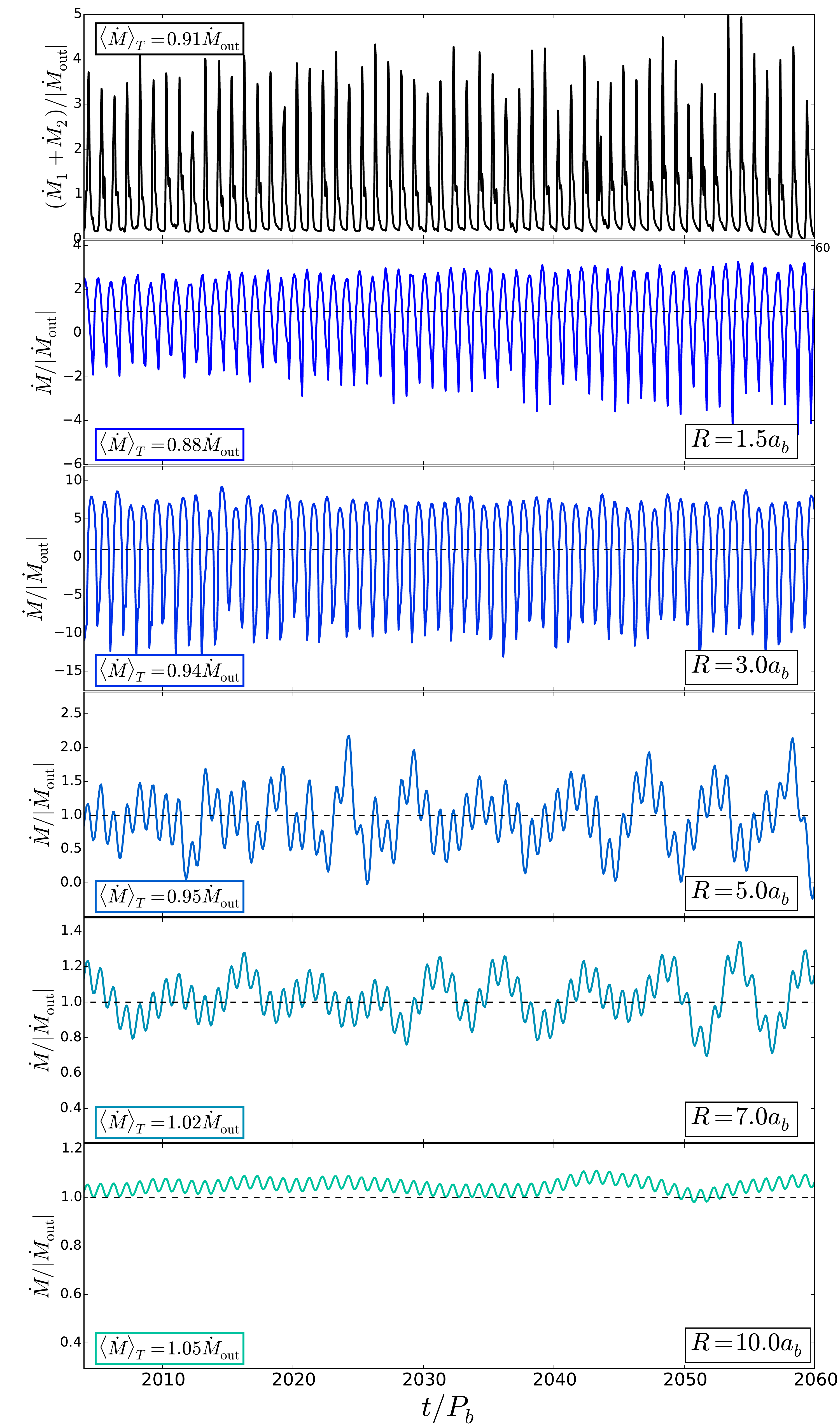}
\vspace{-4pt}
\caption{Accretion rate as a function of time over short time scales ($T=60P_b$) at different radii in the disk. 
From top to bottom, total accretion
rate onto the binary ($\dot{M}_1+\dot{M}_2$) followed by accretion rate ($\dot{M}=-2\pi R\, v_R \Sigma$)   at $R=a_b(1+e_b)$, $3a_b$, $5a_b$,
$7a_b$ and 10$a_b$. The left panels are for $e_b=0$ and the right panels for $e_b=0.5$.
Values are normalized by $\dot{M}_\mathrm{out}$ (represented by the black dashed line), which is 
obtained by identifying the smallest radius at which
the RMS time variability of $\dot{M}$ is less than $1\%$ (a proxy for axisymmetry) and assigning $\dot{M}_\mathrm{out}=\langle \dot{M}\rangle_T$ at that radius. 
For the $e_b{=}0$ case, this radius is identified to be  $\approx11a_b$, and for the $e_b{=}0.5$ case, it is $\approx15a_b$.
In both cases $\dot{M}_\mathrm{out}\approx1.19\dot{M}_0$. 
Variability  of $\dot M$ in the inner region ($R<3a_b$) has a dominant period of $5P_b$ in the case of $e_b{=}0$ and ${\sim}1 P_b$
 in the case of $e_b{=}0.5$.
The averaged $\dot{M}$ over $T=60P_b$ is close to but not exactly $\dot{M}_\mathrm{out}$,
implying that there is variability on timescales longer than shown here. 
Note that accretion rates at a distance of $a_b(1+e_b)$ -- where
an open boundary condition would be imposed in polar-grid simulations -- are always positive for $e_b{=}0$ but 
alternate sign when $e_b{=}0.5$; the latter that cannot be captured by an open boundary. 
This is in consistency with the appearance of shocks inside the cavity (Fig.~\ref{fig:cavity_density})
as a result of the convergence of inflowing and outflowing streams.
\label{fig:accretion_rates_disk}}
\end{figure*}

\begin{figure*}[ht!]
\centering
\includegraphics[width=0.46\textwidth]{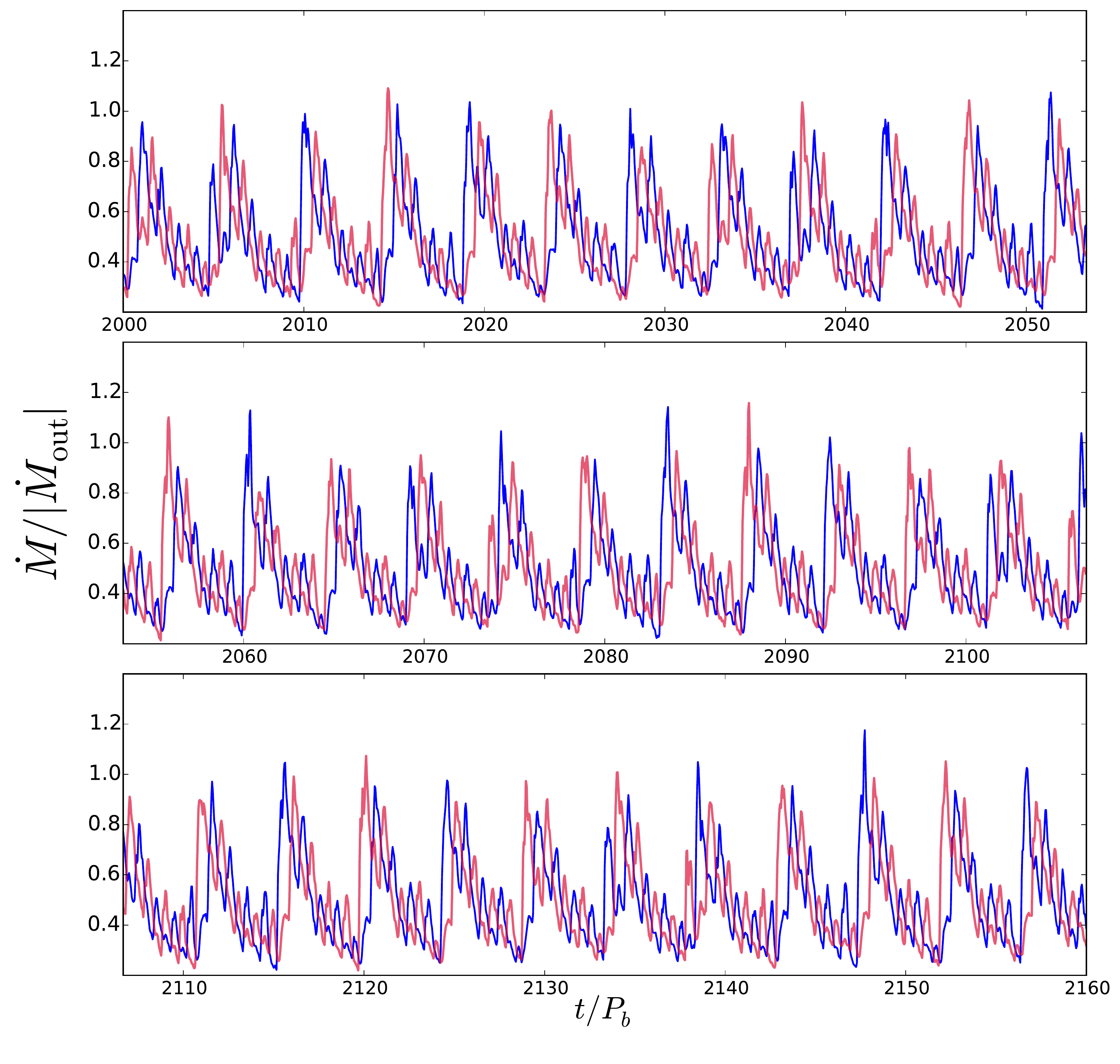}
\includegraphics[width=0.46\textwidth]{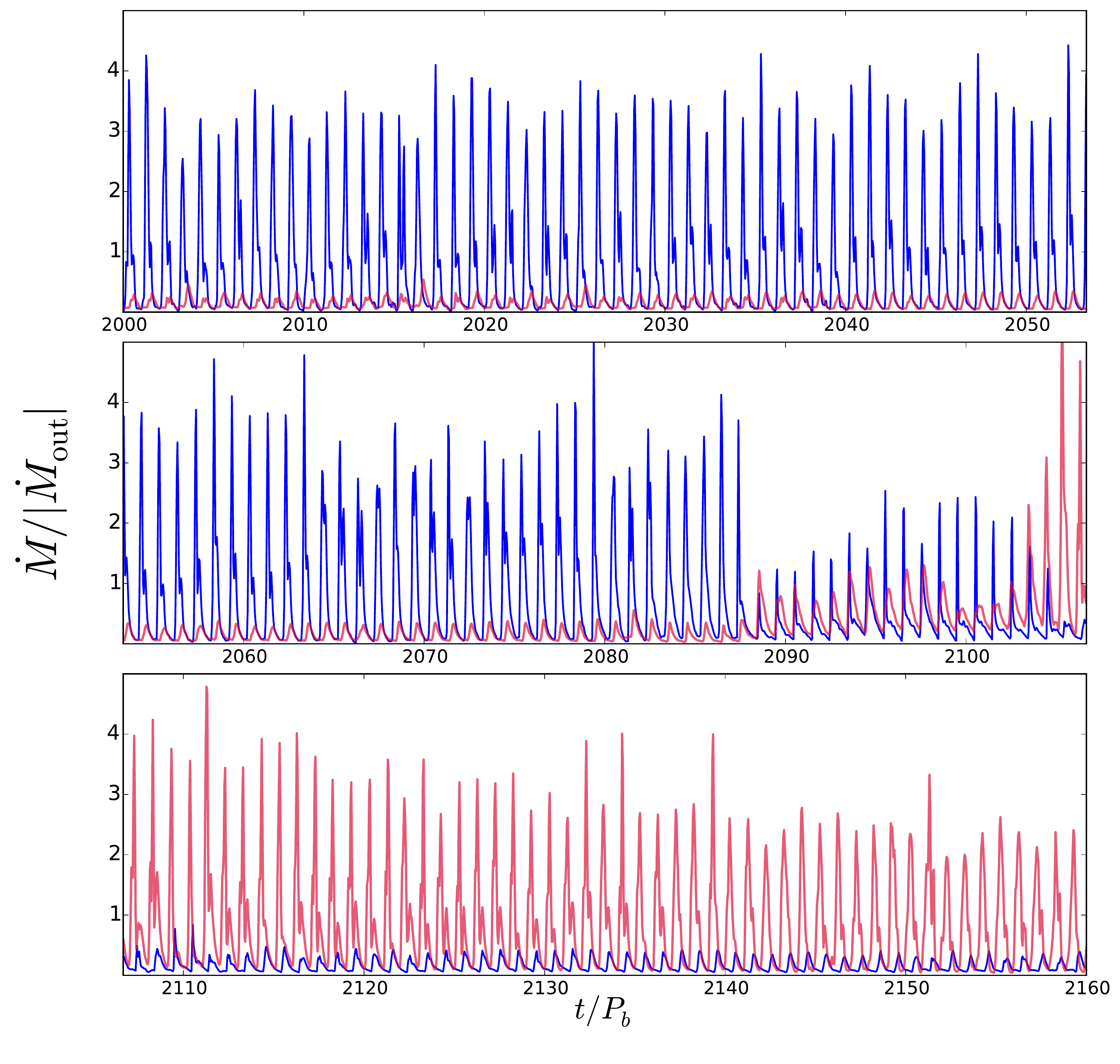}
\vspace{-12pt}
\caption{Accretion rate onto the primary and secondary members of the binary,  $\dot{M}_1$ and  $\dot{M}_2$ in blue and red respectively,
for eccentricities $e_b{=}0$ (left) and $e_b{=}0.5$ (right). Each column spans the range from $2000$ to $2160P_b$ out of a longer
integration period ending at $2600P_b$.  Values are normalized by $\dot{M}_\mathrm{out}$.
 The eccentric binary shows a strong modulation of its accretion rate at the binary orbital period, although long-term trends are also present. The circular 
binary shows strong modulation at both ${\sim}1P_b$ and ${\sim}5P_b$ (as observed also by \citealp{dor13} and \citealp{far14}) and longer-term trends are less
obvious than in the eccentric case. The eccentric binary
experiences a symmetry breaking, with one star accreting between 10 and 20 times more mass than its companion. This trend, however,
is reversed at $t{\sim}2100P_b$ and then reversed back at $t{\sim}2300P_b$ (not shown).
 In the long term (600 orbits), the average accretion rates are $\langle\dot{M}_1\rangle_{600}=0.5\dot{M}_\mathrm{out}$
and $\langle\dot{M}_2\rangle_{600}=0.5\dot{M}_\mathrm{out}$ for the circular binary, and 
$\langle\dot{M}_1\rangle_{600}=0.48\dot{M}_\mathrm{out}$
and $\langle\dot{M}_2\rangle_{600}=0.51\dot{M}_\mathrm{out}$ for the eccentric binary.
\label{fig:accretion_rates_individual}}
\end{figure*}

The initial condition is completed by specifying a rotation curve
\begin{equation}\label{eq:vel_profile}
\Omega^2=\frac{\mathcal{G}M_b}{R^3}\left[1+\frac{3}{4}\left(\frac{a_b}{R}\right)^2\frac{q_b}{(1+q_b)^2}
\left(1+\frac{3}{2}e_b^2\right)\right]
+\frac{1}{R\Sigma}\frac{dP}{dR}~,
\end{equation}
which includes the quadrupole component of the potential and the contribution of the pressure gradient,
and by specifying a radial velocity profile $v_R(R)$. Assuming a standard thin accretion disk,
we impose
\begin{equation}
v_R = \frac{1}{R\Sigma}\frac{\partial}{\partial R}\left(\nu \Sigma R^3 \frac{d\Omega}{dR}\right)
\left[\frac{d}{d R}(R^2\Omega)\right]^{-1}~,
\end{equation}
which in turn specifies the accretion rate profile
\begin{equation}\label{eq:accretion_rate}
\dot{M}(R)=-2\pi R\, v_R(R)\,\Sigma(R)~~.
\end{equation}
Note that this initial $\dot{M}(R)$ starts converging toward $\dot{M}_0$ {\it only beyond} $R\gtrsim 20 a_b$
(at $t{=}0$, $\dot{M}$ equals $1.41\dot{M}_0$ and $1.08\dot{M}_0$ at $R{=}10a_b$ and $15a_b$ respectively),
and thus the disk is not started with a strictly steady accretion profile. 
Unless stated otherwise, 
we initially evolve the system for $2000P_b$ (where
$P_b$ is the binary orbital period) ,
and study the subsequent evolution for an additional 600 binary orbits. This
initial integration time corresponds to two viscous times at
$6a_b$ or ten at  $2a_b$, where the viscous time $t_\nu$ is defined 
for $\nu\propto R^{1/2}$ \citep{lyn74} as
\begin{equation}\label{eq:viscous_time}
t_\nu = \frac{4}{9}\frac{R^2}{\nu}=\frac{2P_b}{9\pi \alpha h_0^2}\left(\frac{R}{a_b}\right)^{3/2}~.
\end{equation}
For $1950$ orbits, we evolve the disk using an open (diode-like) boundary on a set
of controlled cells placed on a ring at $R_\mathrm{in}=a_b(1+e_b)$\footnote{
An open boundary consists of a collection of cells with a prescribed motion playing the role of ``ghost cells''
with outflow boundary conditions \citep{mun13a,mun14}.}. 
 At $t=1950P_b$, boundary cells are ``released'', allowing them to fill in the cavity and form accretion disks around the
individual stars. At $t\gtrsim 2000P_b$,we expect the CBD within $5a_b$ to be (on average) fully relaxed.
We aim to reach a ``relaxed state'' inside the cavity as well, for which
 $\langle \dot{M}_1{+}\dot{M}_2\rangle_T{\approx}\langle \dot{M}(R)\rangle_T{\approx}$constant
 for a wide range in $R$, after averaging over some time interval $T$.
The outer disk($R\gtrsim40a_b$) is in steady-state by construction.
However, there is an intermediate region, with $t_\nu\gg2000P_b$,
that has had no time yet to relax (see below).

\subsection{Long-term Disk Relaxation}\label{sec:relaxation}
Inspection of the initial condition reveals
 that $\dot{M}(R,t{=}0)$ coincides with $\dot{M}_0$
to within $1\%$ only for $R>42a_b$. This is  just
an artifact of the initial condition.
To guarantee $\dot{M}(R)\approx\dot{M}_0$
across all radii, we would need to evolve the system 
for $\sim20000\,P_b$ (or $t_\nu$ at $40a_b$; Eq.~\ref{eq:viscous_time}),
a daunting task for
the simulation work presented here. Instead, after some integration time $t_\mathrm{int}$,
the system has only reached relaxation
within a ``relaxation radius'' 
$R_\mathrm{rel}\equiv a_b\large[\,(9/2)\pi \alpha h_0^2 (t_\mathrm{int}/P_b)\,\large]^{2/3}$
(from setting $t_\mathrm{int}=t_v(R_\mathrm{rel})$
in Eq.~\ref{eq:viscous_time}; see \citealp{raf16}). After
$t=2000P_b$, $R_\mathrm{rel}\approx 9.3a_b$, which is $\ll 42 a_b$,
but sufficiently large for the disk to be nearly axisymmetric outside that radius.
Without reaching global relaxation, we have found that the disk within 
$R_\mathrm{rel}$ receives a gas supply from the partially relaxed portion
of the disk ($R_\mathrm{rel}\lesssim R\lesssim 2R_\mathrm{rel}$)
at a rate $\dot{M}_\mathrm{out}$ slightly larger (by about $10\%$) 
than $\dot{M}_0$, as a result of the initial condition\footnote{
The annulus that supplies gas to the inner CBD can be roughly
estimated by integrating the traveled distance from $R_\mathrm{out,eff}$ 
down to a  $R_\mathrm{rel}$ at a velocity of 
$v_R\sim -\tfrac{3}{2}\nu/R=-\tfrac{3}{2}\alpha h^2 a_b\Omega_b(R/a_b)^{-1/2}$,
giving $(R_\mathrm{out,eff}/a_b)^{3/2}=9\pi\alpha h_0^2 (t_\mathrm{int}/P_b)$
or $R_\mathrm{out,eff}=2^{2/3}R_\mathrm{rel}\approx14.7a_b$ if $t_\mathrm{int}=2000P_b$.
From Eqs.~(\ref{eq:surface_density}) and~(\ref{eq:accretion_rate}), we have 
$\dot{M}(R=14.7a_b,t=0)\approx1.1\dot{M}_0$.}. 
This accretion ``excess" cannot not be removed
with only $2000$ orbits of integration time.

\section{SIMULATION RESULTS}
\subsection{Accretion Flows in the Circumbinary Cavity}
Figure~\ref{fig:cavity_density} shows
several snapshots of the density field for accretion onto a circular
binary (top) and an eccentric one (bottom). The tidal ``cavity'' around the binary
is asymmetric and non-circular, making it difficult to identify an unambiguous
cavity ``radius''. Within the cavity, flow is complex and transient, dominated
by accretion streams, which show significantly more structure
(and unsteadiness) in the eccentric case. In the circular case, the ``streamers''
rotate with the binary, while this is not the case when $e_b{\neq}0$. Similarly,
the cavity shape and contrast nearly repeats itself every half orbit when $e_b{=}0$,
while no such symmetry is observed when $e_b{=}0.5$. 

The symmetries of the $e_b{=}0$ case are also evident from the properties of
the circum-single disks (CSDs):  both CSDs are similar in size, density
and morphology (including $m=2$ spiral patterns in each).  By contrast, the CSDs in the $e_b{=}0.5$ case
show differences in surface density, implying that the members of the equal-mass binary accrete
at different rates. We will further address this ``disk disparity'' in Section~\ref{sec:individual_accretion} below.

\subsection{Accretion rates}\label{sec:accretion_rates}
A schematic description of the circumbinary accretion process is the following:
there are three accretion disks --  the CBD and the two CSDs --  that evolve viscously,  but are connected
via fast accretion bursts owing to the (unstable) tidal streams launched at the inner edge of the CBD. 
If $e_b\neq0$, the outer edges of the CSDs collect new material at apocenter, when the accreting masses are closest
to the CBD inner edge. Subsequently, the incoming material is viscously transported inward
within the CSD, eventually accreting onto the stars at some later phase in the orbit.
The (presumably) much slower rate at which material is transported onto the accreting objects relative to fast 
deposition of material to the outer edge of the CSDs turns the CSDs into ``buffers''. The buffers damp
 the fast oscillations in $\dot{M}$ present in the circumbinary cavity before they reach the stars.

Although compelling, this idealized depiction is clearly too simplistic in the light of the simulations results
of Fig.~\ref{fig:cavity_density}. We compute $\dot{M}(R)$ (Eq.~\ref{eq:accretion_rate}) at different radii\footnote{
We compute $\dot{M}(R,\phi)=2\pi R\, v_R(R,\phi)\,\Sigma(R,\phi)$ for all Voronoi cells 
in the vicinity of radius $R$ and then take an azimuthal average.} 
of the CBD. Figure~\ref{fig:accretion_rates_disk} 
shows $\dot{M}(R)$ at $R=(1+e_b),\,3,5,7$ and $10\times a_b$ as a function of time
for $e_b{=}0$ (left) and $e_b{=}0.5$ (right). In particular, $R=a_b(1+e_b)$ 
is where the innermost boundary would be located in a polar-grid simulation.
The total accretion onto the central masses
$\dot{M}_\mathrm{bin}\equiv\dot{M}_1+\dot{M}_2$ is shown on top. Accretion rates are normalized to a reference value
$\dot{M}_\mathrm{out}$, which is the accretion rate at a radius
where the disk becomes axisymmetric. We measure the mean $\langle\dot{M}\rangle_T$
over a period $T=60P_b$ in each panel (see figure caption).
In the $e_b{=}0$ case (Fig.~\ref{fig:accretion_rates_disk}, left panels), there is a clear accretion modulation 
with period ${\sim}5 P_b$ (roughly the Keplerian period at $R=3a_b$), observed in $\dot{M}_\mathrm{bin}$
(top panel) as well as in the CBD out to $R=7a_b$.
Right at the putative edge of the CBD disk (${\approx}3a_b$), the evolution
of $\dot{M}$ turns significantly more complex (a quasi-periodicity of $5P_b$
is still present), and extremely variable
in amplitude (going from $-40$ to $+20$ $\dot{M}_\mathrm{out}$).
For $R{<}3a_b$, modulation of $\dot{M}$ is dominated by a $1P_b$ component
superposed to major bursts that repeat every ${\sim}5 P_b$.
This bursty accretion has been seen
in previous numerical experiments \citep{dor13,far14}, being attributed to an over-dense ``lump'' that forms
at the rim of the cavity, and gets periodically ``flung'' onto the binary (see Fig.~\ref{fig:cavity_density}, top panels).
Note that at $R{=}1a_b$ -- where a polar-grid code would place the outflow computational boundary -- 
$\dot{M}$ is always positive, and thus artifacts introduced by diode-like boundary conditions (not allowing
for material with $v_R>0$ to enter the domain) are minimal.
The two top panels of Fig.~\ref{fig:accretion_rates_disk} show good qualitative agreement with each other,
the differences being (1) a delay in the time of the accretion burst to reach the stars, and (2) a reduction
of the amplitude of the variability; these differences are consistent
with the buffering nature of a viscous disk.

The $e_b{=}0.5$ case (Fig.~\ref{fig:accretion_rates_disk}, right panels) shows
much more complex $\dot{M}$ variability outside
$R{=}3a_b$. By contrast, for $R{<}3a_b$, variability seems simpler than around a circular binary.
In particular, ${\sim}1P_b$ is the dominant modulation period, although trends of periods longer than 
$60P_b$ are also noticeable. The amplitude of the oscillations in $\dot{M}_\mathrm{bin}$
(Fig.~\ref{fig:accretion_rates_disk}, top right panel) can be as high as $5 M_\mathrm{out}$
in contrast with ${\sim}1.5 M_\mathrm{out}$ for the $e_b{=}0$ case. Another striking difference from
the $e_b{=}0$ case is the value of $\dot{M}$ around the binary. The imaginary boundary
at $R{=}a_b(1+e_b)$ (second right panel from top) shows alternating negative and positive values in
$\dot{M}$. Evidently, a diode-like boundary placed at $R{=}a_b(1+e_b)$ could not capture this sign-changing
behavior.

\subsection{Individual Accretion Rates and Correlation with Periastron Separation}\label{sec:individual_accretion}
In Fig.~\ref{fig:accretion_rates_individual} we show accretion onto the individuals stars
$\dot{M}_1$ and $\dot{M}_2$ over 160 binary orbits. For the $e_b{=}0$ case (left panels), 
the symmetry between the primary and secondary 
is remarkable, as it is expected for $q_b=1$. This is at odds with the results
of \citet{far14} which shows a mild ``symmetry breaking'' in $\dot{M}_i$. Both
$\dot{M}_1$ and $\dot{M}_2$ show the bursty nature of the two top left panels of Fig.~\ref{fig:accretion_rates_disk},
although they do not perfectly lie on top of each other; instead, one star undergoes
an accretion burst before its companion. The lag between the two bursts
 is about a half orbit, although the sign of the lag alternates on each major burst.
\begin{figure}[hb!]
\includegraphics[width=0.5\textwidth]{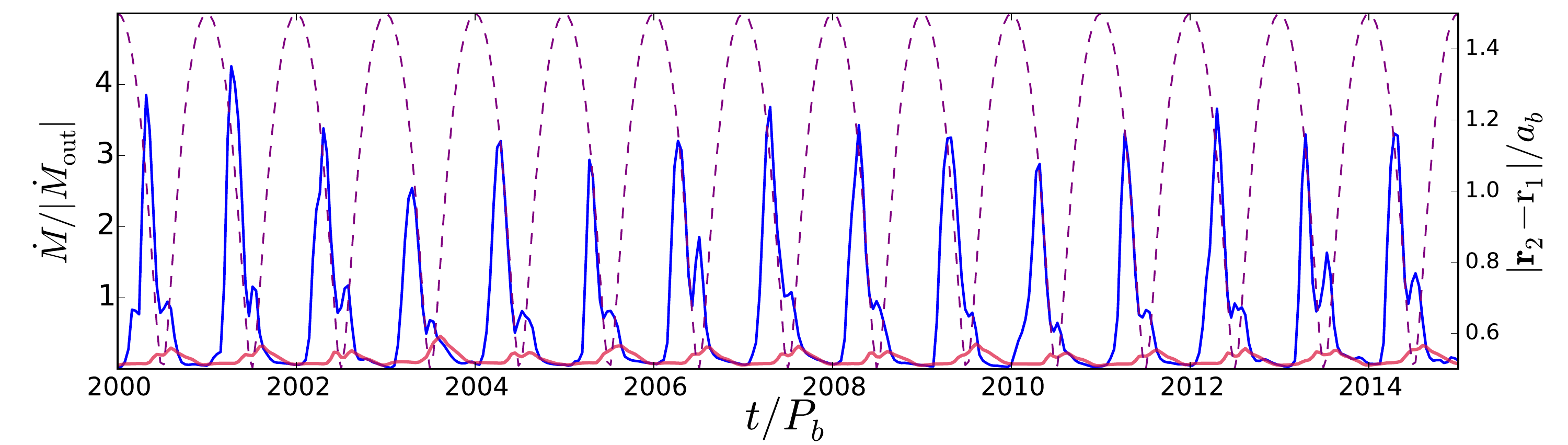}
\caption{Individual accretion rates for the primary (blue) and secondary (red) in the $e_b{=}0.5$ case over an interval of 15 orbits contrasted to 
orbital separation (purple dashed curve). The primary (only temporarily dominating the accretion rate, Fig.~\ref{fig:accretion_rates_individual}) shows a double peaked 
behavior, with a dominant peak consistently located at an orbital phase of $\phi=-0.2$ (relative to time of pericenter) and a minor peak immediately after/during
pericenter. The secondary also shows a double peaked accretion curve (immediately before and after pericenter), although with comparable amplitudes.
\label{fig:accretion_rates_separation}}
\end{figure}
\begin{figure*}[t!]
\centering
\includegraphics[width=0.73\textwidth]{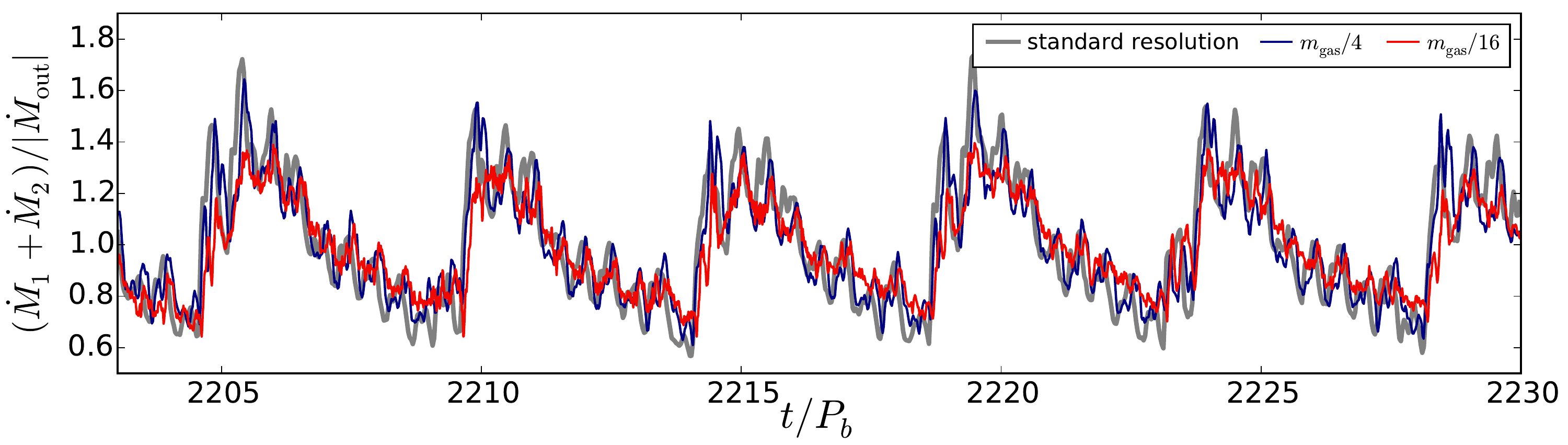}
\includegraphics[width=0.74\textwidth]{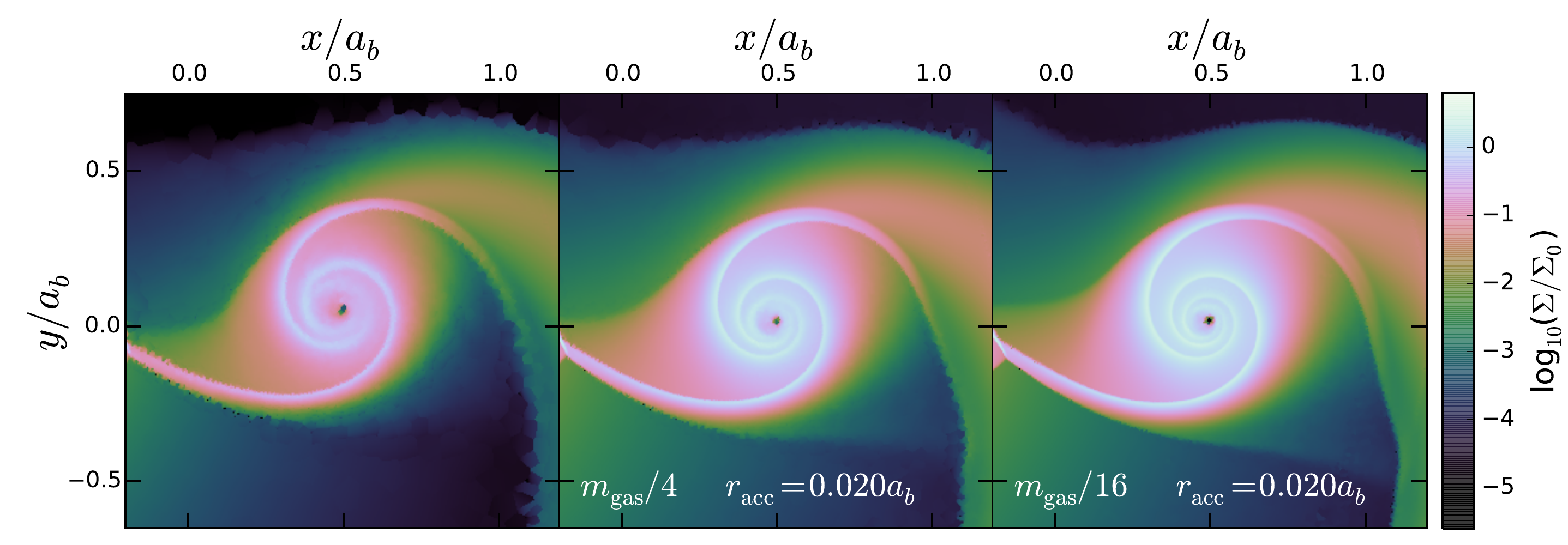}
\vspace{-6pt}
\caption{Top panels: density field around the primary at $t=2220P_b$
for a resolution mass of $m_\mathrm{gas}=6.3\times10^{-7}\Sigma_0 a_b^2$ (left),
$m_\mathrm{gas}/4$ (center) and $m_\mathrm{gas}/16$ (right). In all cases $r_\mathrm{acc}=0.02a_b$
(Section~\ref{sec:resolution}).
Bottom panel: stellar accretion $\dot{M}_\mathrm{bin}$ as a function of time for the three different
mass resolutions. Both the burst in accretion with period $\sim5P_b$ and the high-frequency modulations
with periods $\lesssim1P_b$ are prevalent at all different resolutions, strengthening the hypothesis that
the prominent $m{=}2$ features drive the accretion onto the stars.
When the same experiment is run after decreasing $r_\mathrm{acc}=0.02a_b$ and $s$ (by a factor of 2 and then by a factor of 4)
in addition to $m_\mathrm{gas}$ (by a factor of  4 and 16 respectively), the long term modulation of the accretion rate persists, but the high frequency component is
progressively washed away. This, again, is consistent with transport via resonant torques, as for $r_\mathrm{acc}$ sufficiently
small, the tidal forcing from the secondary becomes negligible (Eq.~\ref{eq:tidal_potential}).
\label{fig:resolution_accretion}}
\vspace{-3pt}
\end{figure*}

For $e_b{=}0.5$ (right panels of Fig.~\ref{fig:accretion_rates_individual}),
the symmetry breaking between $\dot{M}_1$ and $\dot{M}_2$ is evident.
Despite the having $q_b=1$,
over the first $\sim90$ orbits, $\dot{M}_1$(blue) is 10-20 times larger than $\dot{M}_2$ (red). Interestingly,
after 100 orbits, this behavior switches to $\dot{M}_2\gg\dot{M}_1$, only to switch back to 
$\dot{M}_1\gg\dot{M}_2$ at $t=2300 P_b$ (not shown). Over timescales of $\sim600P_b$, we have that
$\dot{M}_1\approx\dot{M}_2$, recovering -- in a time-averaged sense -- the symmetry that is to be expected when $q_b=1$.
The reason for this dramatic difference between $\dot{M}_1$ and $\dot{M}_2$ must originate in a symmetry breaking
in the CBD itself. If the CBD is eccentric, the relative longitude of pericenter $\varomega_d-\varomega_b$\ (where $\varomega_d(R)$
specifies the orientation of a given elliptical portion of the CBD disk) will determine the timing of mass transfer from the CBD to the binary.
In principle, an eccentric disk should precess around the binary at a rate $\dot{\varomega}_d$,
implying that, if one of the accreting objects is benefited by an increased $\dot{M}$ at any given time,
at some later time preferential accretion should
be reversed.
A relevant precession
rate is that of the inner rim of the CBD, at a radius of $R_\mathrm{cav}\sim 2-3 a_b$. In the 
limit of a pressure-less particle disk, the precession rate around an eccentric binary is
\begin{equation}
\begin{split}
\dot{\varomega}_d&{\simeq}~{\frac{3\Omega_b}{4}}\frac{q_b}{(1+q_b)^2}
{\left(1{+}\frac{3}{2}e_b^2\right)}{\left(\frac{a_b}{R}\right)^{7/2}}
{\sim}{~}0.006\,\Omega_b\left(\frac{3a_b}{R}\right)^{7/2},
\end{split}
\end{equation}
which corresponds to precession period of a few hundred $P_b$ at $R\sim3a_b$. This precession period roughly coincides with
the period of alternation of dominant accretion shown in Fig.~\ref{fig:accretion_rates_individual}. Future work
will take deeper look into the properties of precessing eccentric CBDs (Miranda, Mu\~noz \& Lai, in prep).

In Fig.~\ref{fig:accretion_rates_separation}, we show a portion of Fig.~\ref{fig:accretion_rates_individual} ($e_b{=}0.5$ case, right panels)
overlaid with the binary separation $|\mathbf{r}_2-\mathbf{r}_1|$. In the case where
$\dot{M}_\mathrm{bin}$ is dominated by $\dot{M}_1$ (blue curve) accretion peaks before pericenter passage, exhibiting a
minor second peak exactly at pericenter. This is in partial agreement the simulation results of \citet{gun02} and \citet{val11},
although the accretion burst peaks noticeably before pericenter, and the burst duration spans a significant fraction of the orbital period.

\subsection{Tidal Torques and the Effect of Increased Resolution}\label{sec:resolution}
We now examine the buffering nature of the CSD discussed above (Section~\ref{sec:accretion_rates}).
When the accretion radius $r_\mathrm{acc}$ (Section~\ref{sec:arepo}) is
much smaller than the size of the CSD $R_\mathrm{cs1}$, the accretion time $t_\mathrm{acc,cs1}$ within a CSD 
is roughly the viscous time at the disk edge\footnote{
If the condition $r_\mathrm{acc}\ll R_\mathrm{cs1}$ is not satisfied, a more general expression for the accretion time is
$t_\mathrm{acc,cs1}=t_{\nu,\mathrm{cs1}}[1-({r_\mathrm{acc}}/{R_\mathrm{cs1}})^{3/2}]$.
} 
$t_{\nu,\mathrm{cs1}}={2P_b}/({9\pi\alpha h_0^2})~({R_\mathrm{cs1}}/{a_b})^{3/2}\sqrt{1+q_b}$. This timescale 
enables the damping of fast modulations and sets a delay between the time of gas deposition onto the CSD and the time of actual
accretion onto the stars. With an estimate of $R_\mathrm{cs1}$\footnote{
We replace $R_\mathrm{cs1}$ with the Eggleton approximation of the Roche radius \citep{egg83}: 
\begin{equation}
\big(R_\mathrm{cs1}/a_b\big)=0.49q_b^{-2/3}\Big/\left[{0.6q_b^{-2/3}+\ln\left(1+q_b^{-1/3}\right)}\right]~.
\end{equation}
}, 
we have  $t_\mathrm{\nu,cs1}{\approx}23P_b$. This time
may be short enough to enable the accretion burst of period ${\sim}5P_b$ to reach the stars, but it is perhaps too long
to allow for the persistence of the high frequency oscillations (periods ${\sim}1P_b$ and shorter; 
Figs.~\ref{fig:accretion_rates_disk} and~\ref{fig:accretion_rates_individual}, left panels).
However, the fast oscillations in  $\dot{M}_\mathrm{bin}$ might be explained
by mass transport via tidal torques. In a circum-primary frame (primed coordinates), the potential on the CSD
 due to the secondary is (e.g., \citealp{mir15})
\begin{equation}\label{eq:tidal_potential}
\begin{split}
\Phi_\mathrm{sec}(\mathbf{r}')
&\approx \mathcal{G}M_b \frac{q_b}{1+q_b}\frac{1}{4}\left(\frac{r'}{a_b}\right)^2\big[1+3\cos (2\Omega_b\,t)\big]
\end{split}
\end{equation}
where we have dropped a constant term. The time-dependent term can excite $m{=}2$ modes
in the density field of the circum-primary disk (Fig.~\ref{fig:cavity_density}), which may explain
the high-frequency modulation of $\dot{M}_\mathrm{bin}$ as the spiral overdensities move into the accretion region.
Note that our accretion routine is somewhat resolution-dependent (cells are drained depending on their location, regardless
of their total mass content; see \citealp{mun15}). For a given $\dot{M}_0$,
the average number of cells being accreted in an interval $\Delta t$ is $N_\mathrm{acc}=(\dot{M}_0/m_\mathrm{gas})\Delta t$,
with a ``signal-to-noise ratio"\footnote{
The imposed accretion rate $\dot{M}_0$ and mass resolution are related by 
$\dot{M}_0=3\sqrt{5}\pi\alpha h_0^2(R_{\mathrm{cav},0}/5a_b)^{1/2}[m_\mathrm{gas}\Omega_b/(6.7\times10^{-3})]$~.
} 
of $\sqrt{N_\mathrm{acc}}\approx200\sqrt{\Omega_b\Delta t}$. For  
$\Delta t\sim 0.05 P_b$, the uncertainty in the measured $\dot{M}_\mathrm{bin}$ is $\sim1\%$, small enough to be confident in
the general features of Fig.~\ref{fig:accretion_rates_individual}, but large enough to justify a convergence study of the high-frequency
modulations \citep[see][]{pak16}. Fig.~\ref{fig:resolution_accretion} (top panels) shows the circum-primary density field at three different gas resolutions
 (while keeping $r_\mathrm{acc}$ and the softening length $s$ fixed),
confirming the prevalence of $m{=}2$ spiral arms. The bottom panel of Fig.~\ref{fig:resolution_accretion}  
shows $\dot{M}_\mathrm{bin}$ at the same three resolutions, 
confirming the major accretion burst, and that the rapid oscillations are real and likely a result of a time-dependent
forcing (Eq.~\ref{eq:tidal_potential}). Note that the strength of this forcing in Eq.~\ref{eq:tidal_potential}) is negligible for $r'\ll a_b$; thus, one can expect the influence
of the companion (and thus the high frequency modes of $\dot{M}_\mathrm{bin}$) to be drastically reduced as $r_\mathrm{acc}$ is made smaller.
We repeat the resolution experiments of Fig.~\ref{fig:resolution_accretion} (not shown), this time decreasing $r_\mathrm{acc}$ and $s$ in addition to $m_\mathrm{gas}$.
We find that the fast modulations are progressively damped out; and for very small $r_\mathrm{acc}$,
 only the major accretion bump survives, as the accretion (buffering) time
within the CSD is not long enough to entirely erase it.

\section{Summary and implications}\label{sec:summary}

We have presented two-dimensional, viscous flow simulations of circumbinary disk
accretion for the first time using a finite-volume method on a freely
moving Voronoi mesh.  Previous simulations based on structured moving
grids were restricted to circular binaries \citep[e.g.][]{far14}.
Using {\footnotesize AREPO},
we can robustly simulate accretion onto arbitrarily eccentric binaries, without
the constraints imposed by structured grids.  In our simulations, we are
able to follow the mass accretion through a wide radial extent of the
circumbinary disk, leading to accretion onto individual members of the
binary via circumstellar disks.

Our simulations have revealed dramatic differences between the accretion
behavior of circular and eccentric binaries:

(1) In agreement with previous studies \citep[e.g.][]{dor13,far14}, we find that
accretion onto equal-mass, circular binaries exhibits quasi-periodic
variabilities with a dominant period of ${\sim}5P_b$ (where $P_b$ is
the binary period), corresponding to the orbital period of the
innermost region of the circumbinary disk.  By contrast, accretion
onto eccentric binaries displays larger-amplitude variabilities
dominated by pulses with periods of ${\sim}1\,P_b$ (see Fig.~2).

(2) For equal-mass circular binaries, we find that the accretion rates
onto individual stars are quite similar to each other, following an essentially
identical accretion pattern in time
(Fig.~\ref{fig:accretion_rates_individual}, left panels).  This result
differs from the simulations by \citet{far14}, which produced an
appreciable disparity between the individual stellar accretion rates.
By contrast, we find that accretion onto eccentric binaries exhibits
strong symmetry breaking: for a period of time lasting $\sim 200\,P_b$
(which corresponds to the apsidal precession period of the innermost region of the
circumbinary disk), one of the stars can accrete 10 to 20 times more
than the companion (Fig.~\ref{fig:accretion_rates_individual}, right
panels).  This disparity alternates over timescales of ${\sim} 200
P_b$, such that the long-term accreted masses onto individual stars
are the same.

In addition to using a novel moving mesh code ({\footnotesize AREPO}) that resolves
the binary-disk system over a large dynamical range, an important
feature of our study is that we carry out our simulations for a
sufficiently long time (thousands of binary orbits) and with a proper
initial setup. The inner circumbinary disk and the individual
circumstellar disks reach a quasi-steady state in which
the time-integrated mass accretion is the same
across different regions of the system.
The ability to reach quasi-steady state gives us confidence
that the pulsed accretion behavior uncovered in this
paper is not the result of artificial initial conditions.

Our results can be compared to the observations of pulsed accretion in
binary T-Tauri stars (BTTS) \citep{jen07,muz13,bary14} and can shed
light on the origin of the quasi-periodic variability (in broadband
photometry and near-IR line fluxes) observed in these systems.
In our simulations, accretion onto eccentric binaries peaks before and
during pericenter 
but never at apocenter.  
This appears to contradict with the observation of 
BTTS DQ~Tau (with $e_b=0.56$), which 
exhibits flaring events during apocenter \citep{bary14}.
This apparent discrepancy can be easily understood 
once we recognize that the size of the accretion
region $r_\mathrm{acc}$ can strongly affects the measured variability of
accretion rates (Section~\ref{sec:resolution}). In the case of DQ~Tau, 
pericenter passage of the binary (at the separation of
$r_p=a_b(1-e_b)\approx0.05$~AU; \citealp{mat97}) would limit the size
of circumstellar disks to be less than ${\sim}r_p/3\approx0.02$~AU,
i.e., about 2-3 pre-main-sequence stellar radii, making circumstellar
disk accretion irrelevant (especially if the stellar magnetospheres
are indeed colliding at periastron; \citealp{sal10}), with accretion
proceeding almost directly from the streamers to the stars.  In
addition, it is possible that the shocked gas responsible for line
emission is not strictly confined to the stellar photospheres
\citep{cal98}, but located elsewhere in the circumbinary cavity
\citep[e.g.][]{bary14}. Indeed, our eccentric binary simulations do
show that shocks appear as material is swung out from the edges of the
circumstellar disk at each close passage; this outflowing material
meets the inflowing accretion streams from the circumbinary disk.  As
the disk flow is highly supersonic, the eccentric accretion streams
could shock against material at a relative Mach number of
$\mathcal{M}\sim10$. We plan to explore the observational signatures
of these shocks in future work.

Finally, although we have focused on accretion onto pre-main sequence
binaries in this paper, our simulations also have implications for
accretion onto supermassive binary black holes (SMBBHs).  In
Section~\ref{sec:resolution} we have discussed how the size of the
accreting region $r_\mathrm{acc}$ can affect the variability of
accretion rates, such that when $r_\mathrm{acc}\rightarrow 0$,
modulations of accretion on timescales much shorter than the circumstellar disk
viscous time are damped out. Since we expect $r_\mathrm{acc}\ll a_b$
for SMBBHs, the individual black holes in a binary would accrete
at the nominal supply rate, suppressing fast variability.  If this is
the case, the mechanism behind the observational hints of photometric
variability of SMBBHs would most likely be due to Doppler beaming,
as suggested by \citet{dor15}, rather than to gas dynamics within the
circumbinary cavity.
 
\acknowledgments{DJM thanks Volker Springel 
for making {\footnotesize AREPO} available for use in this work.
We thank Ryan Miranda for useful discussions.  This work has been
supported in part by NSF grant AST-1211061, and NASA grants NNX14AG94G
and NNX14AP31G.  }

\bibliographystyle{apj}

\end{document}